\documentclass[final,5p,times,twocolumn,authoryear]{elsarticle}

\usepackage{amssymb}
\usepackage{amsmath}
\usepackage{lipsum}
\usepackage{graphicx}
\usepackage{dcolumn}
\usepackage{bm}
\usepackage{soulutf8}
\usepackage[dvipsnames, svgnames]{xcolor}
\usepackage{braket}
\usepackage{color}
\sethlcolor{green}
\colorlet{MySeaGreen}{SeaGreen!30!Honeydew}
\colorlet{MyPeachPuff}{PeachPuff!40}

\journal{Diamond and related materials}

\begin{document}

\begin{frontmatter}



\title{Sensing with near-infrared laser trapped fluorescent nanodiamonds}


\author[1]{Arthur Dervillez}
\author[1,3]{Fatemeh Kalantarifard}
\author[1,2]{Luca Troise}
\author[2]{Alexander Huck}
\author[1]{Kirstine Berg-S\o rensen\fnref{correspondence to: kibs@dtu.dk}}
\affiliation[1]{organization={Department of Health Technology, Technical University of Denmark},
            addressline={}, 
            city={Kongens Lyngby},
            postcode={2800}, 
            state={},
            country={Denmark}}
\affiliation[2]{organization={Department of Physics, Technical University of Denmark},
            addressline={2800}, 
            city={Kongens Lyngby},
            postcode={}, 
            state={},
            country={Denmark}}
\affiliation[3]{organization={Department of Natural and Mathematical Sciences, Ozyegin University},
            addressline={}, 
            city={Istanbul},
            postcode={34794}, 
            state={},
            country={Turkey}}            

\begin{abstract}
Biosensing based on optically trapped fluorescent nanodiamonds potentially allows to resolve biochemical processes inside living cells at a desired intracellular location. Towards this goal, we investigate near infrared (NIR) laser irradiation at 1064 nm on fluorescent nanodiamonds (FNDs) containing nitrogen-vacancy (NV) centers. The 1064 nm NIR wavelength is a popular choice for optical trapping because of its low absorption in bio-samples.  
By conducting comprehensive experiments, we aim to understand if and how NIR exposure influences the fluorescence and sensing capabilities of FNDs and to determine the potential implications for the use of FNDs in various sensing applications. Our experiments exposed FNDs to varying intensities of NIR laser light while carefully monitoring their optical and magnetic properties. Key measurements included all-optical fluorescence relaxation, optical spectroscopy, and optically detected magnetic resonance (ODMR) spectra. The findings reveal how increased NIR laser power correlates with alterations in ODMR central frequency but also that charge state dynamics under NIR irradiation of NV centers play a role. We demonstrate that FND biosensing works well with a protocol involving both NIR and green light, while mitigating the effect of NIR. 
\end{abstract}

\begin{graphicalabstract}
\includegraphics[width=8cm]{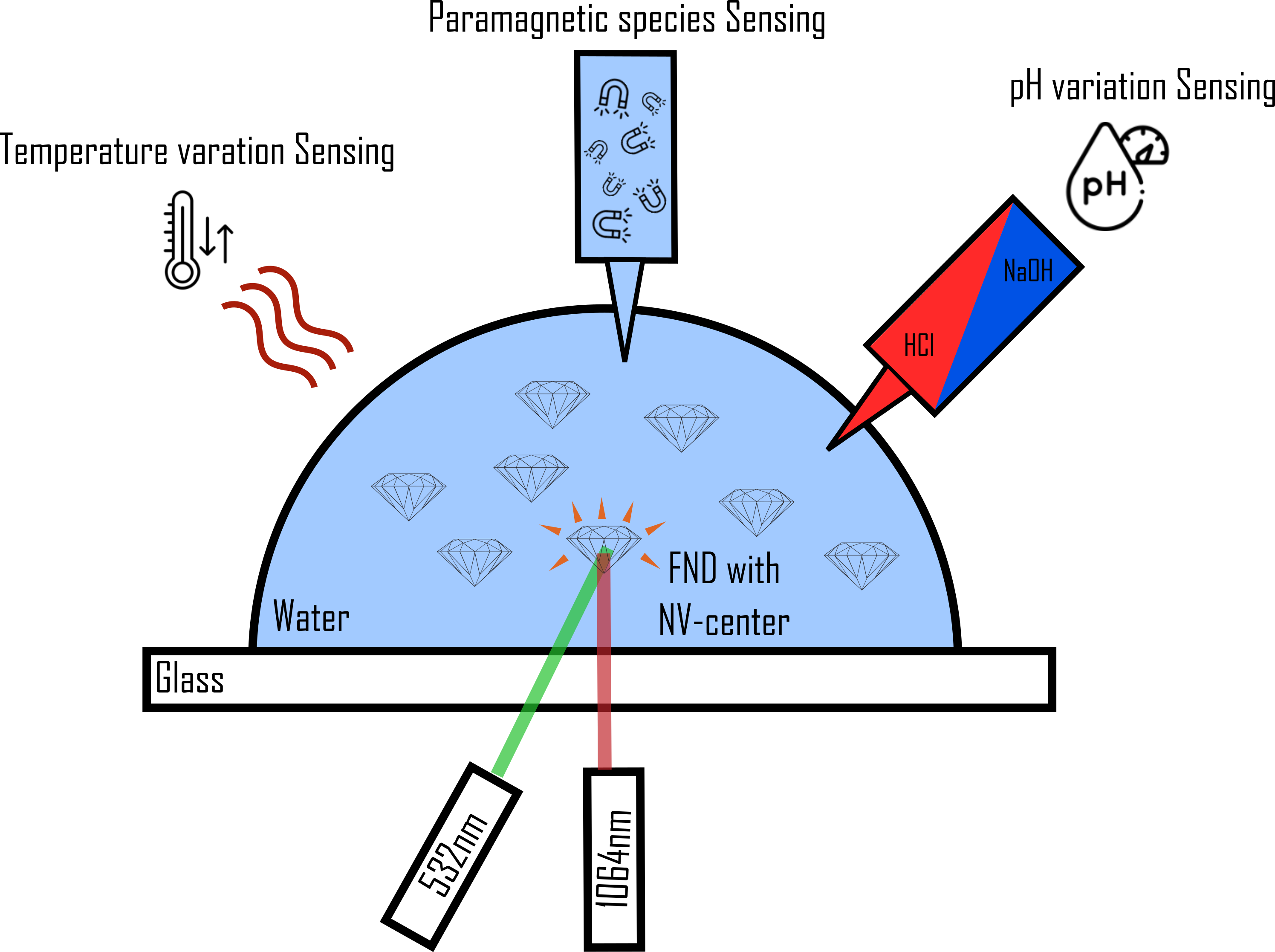}
\end{graphicalabstract}

\begin{highlights}
\item Demonstration of multi-modal NV-center based sensing with optically trapped nanodiamonds 
\item Exploring the effect of a 1064~nm NIR trapping laser on the NV-center readouts
\end{highlights}


\end{frontmatter}


\section{\label{sec:intro} Introduction}

Fluorescent nanodiamonds (FNDs) hosting negatively-charged nitrogen-vacancy (NV\(^{-}\)) centers combine bright, photostable fluorescence with quantum-coherent electron-spin states, making them uniquely suited for bio-sensing and bio-imaging at the nanoscale\,\cite{Schirhagl2014,Hsiao2016}. Because NV spins can be optically polarised and read out at room temperature, changes in their spin-state populations translate directly into optical signals, enabling quantitative sensing of magnetic fields, temperature and local electro-chemical potentials\,\cite{Doherty2013,Rondin2014}. The ultimate performance of those sensors, however, is governed by the spin- and charge-state photodynamics of the NV centers themselves; understanding how an experimental control parameter perturbs those dynamics is therefore a prerequisite for accurate, high-sensitivity measurements.

The negatively charged NV center is an electronic spin-1 system with a ground-state zero-field splitting (ZFS) \(D \approx 2.87 ~\mathrm{GHz}\) that separates the \(m_s = 0\) and \(m_s = \pm 1\) sub-levels. External magnetic fields lift the degeneracy of the \(m_s = \pm1\) levels through the Zeeman effect, while lattice strain, temperature and electric fields shift the ZFS and the optical transition energies via the Stark effect\,\cite{Maze2008,Dolde2011}. In addition, the relative abundance of the neutral (NV\(^0\)) and negative (NV\(^{-}\)) charge states—and thus the fluorescence spectrum and intensity—depends sensitively on the local Fermi level and on photo-ionisation/recombination pathways. These intertwined dependencies let NV centers act as multi-modal nano-probes of their local environment, provided that the separate interaction between all external controls (laser beams, microwave fields, etc.) and the FNDs are themselves well understood and do not compromise sensor performance.

A particularly powerful external control is optical trapping. Near-infra-red (NIR) optical tweezers can confine and manoeuvre sub-100 ~nm diamonds inside living cells or complex micro-fluidic environments, offering spatially resolved sensing with a resolution set by the nanoparticle size rather than by the optical diffraction limit\,\cite{Horowitz2012}. Operating the trap at 1064 ~nm keeps tissue absorption low and mitigates photo-damage\,\cite{Svoboda1994}, yet the same intense NIR field may also heat the nanodiamond, alter NV charge-state kinetics or modify spin-lattice relaxation~\cite{Lai2013,Geiselmann2013,Ji2016}. Earlier work on bulk diamond or levitated nanodiamonds indeed reported substantial NIR-induced changes in fluorescence and the spin relaxation \(T_{\rm relax}\)\,\cite{Neukirch2013,Ji2016}, but a clear picture for FNDs dispersed in aqueous media—arguably the configuration of greatest practical relevance-is missing.

Here, as illustrated by the graphical abstract, 
we systematically address that gap. Using freely diffusing and optically trapped FNDs in water and in biologically relevant buffer solutions, we quantify how a 1064~nm focused trapping beam with an average power between 30–60~mW focused to a spot with an Airy diameter less than 450~nm affects NV optical and spin properties. We employ three complementary read-outs, each indicative of a particular NV property: i) photoluminescence (PL) spectroscopy to monitor charge-state redistribution between NV$^0$ and NV$^-$ \,\cite{Aslam2013,Geiselmann2013,Sow2020}; ii) optically detected magnetic resonance (ODMR) at zero bias field to track shifts in the resonance frequencies with kHz precision which we discuss as a measure of temperature change \,\cite{Gruber1997,Acosta2010,Doherty2014,Hall2016}; and iii) fluorescence relaxometry to probe magnetic-noise spectra—e.g.\ from free radicals or \(\mathrm{Gd}^{3+}\) ions—at the NV transition frequency\,\cite{Tetienne2013,Mzyk2022}. 
Together, these observables allow us to address and potentially disentangle three mechanisms that NIR light has been suggested to induce, namely a) Photothermal heating of the FND, shifting the ZFS via the temperature coefficient \(dD/dT \approx -74~\mathrm{kHz\,K^{-1}}\)\,\cite{Acosta2010,Lai2013}; b) Power-dependent ionisation/recombination that re-weights NV\(^{-}\) and NV\(^0\), thereby changing both ODMR spectral properties and PL brightness\,\cite{Aslam2013,Ji2016}; and c) Modified surface sensitivity, whereby the NIR field—through either heating or altered charge state—amplifies or suppresses the NV response to environmental parameters such as pH or paramagnetic species concentration\,\cite{Fu2010}. 

Initially, we compare these effects in air-dried FNDs, then in trapped FNDs immersed in an aqueous solution. Then, to address potential impact of NIR on sensing capabilities of the FNDs, we measure on trapped FNDs in solutions of controlled pH and temperature, and finally in the presence of a paramagnetic relaxant (\(\mathrm{Gd}^{3+}\)). Thereby, we establish quantitative limits on NIR-induced artefacts and delineate operating regimes where accurate, spatially resolved biosensing is possible.

\section{\label{sec:Methods}Methods} 

\subsection{\label{sec:Methods-FNDprep}Sample Preparation}
Fluorescent FNDs with a nominal diameter of 120 nm (Sigma-Aldrich, 798088) and an NV concentration of 3ppm were used for most of the experiments reported in this article if not stated otherwise. 
To prepare the FND suspension, 1 µL of the commercial stock solution was diluted in 300-700 µL of deionized water depending on the desired concentration and before sonicating for 10 minutes. Next, the petri dish (Mattek, part no P35G-1.5-10-C) was prepared before deposition of FNDs. For this, the petri dish was thoroughly cleaned
and plasma ionised in order to make the surface hydrophilic, ensuring an even distribution of FNDs on the surface. Plasma treatment was applied for 30 seconds in a simple custom-device with a closed chamber in PMMA and under ambient conditions. Plasma was generated with a BD-20AV high-frequency generator (230 V, ETP, IL, USA) with an electrode positioned close to the petri dish surface. 
Approximately 3 µL of the FND suspension was then carefully deposited onto the clean petri dish using a micropipette and the sample was left to dry at room temperature. 
The preparation of FNDs for experiments in the liquid environment also starts with the dilution steps and cleaning of the petri dish, but to maintain the FNDs in solution, the petri dish was not plasma treated and the solution was not left for evaporation.

FNDs with a nominal size of 70~nm were used for sensing experiments as they are preferably used in our most recent cell experiments~\cite{Niora2024,Mzyk2025,Mzyk2026}. The 70~nm FNDs (Sigma-Aldrich 1003626896) were prepared following the same protocol. A suspension of FNDs was created by diluting 1 µL of the commercial stock solution in 1000 µL of deionized water, followed by sonication for 10 minutes to ensure a homogeneous dispersion. Approximately 50 µL of the suspension was deposited in a sterile Petri dish that was freshly unwrapped.

\subsection{\label{sec:Methods-setup}Optical Setup}
\begin{figure}[ht!]
\centering
  \includegraphics[width=0.45\textwidth]{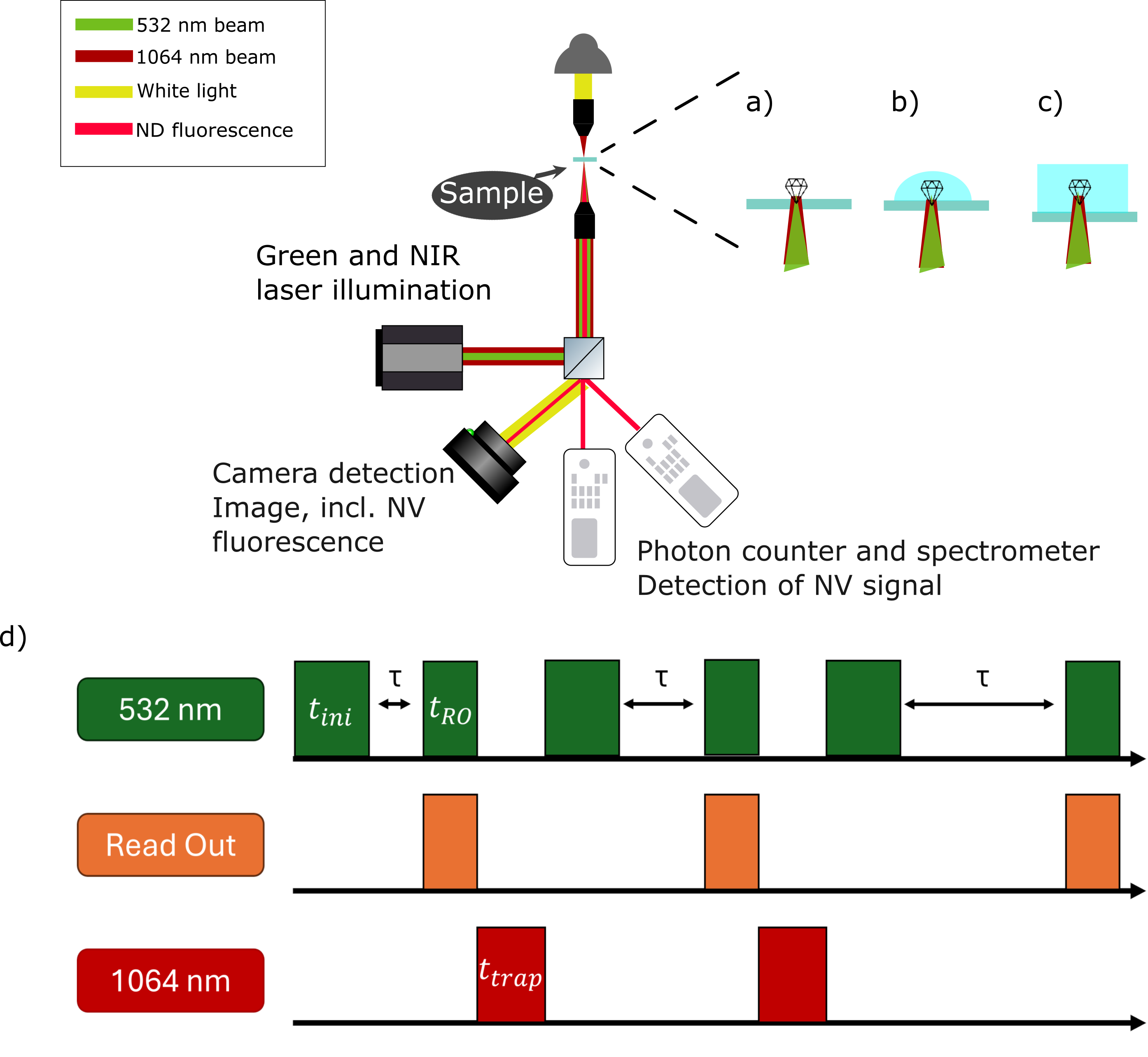}
  \caption{\textbf{Essential elements of the optical setup.} NV centers are excited with 532 nm green light, and nanodiamonds are optically trapped with the 1064 nm NIR light. Detection modalities include bright-field, wide-field imaging on a camera as well as photon counting and fluorescence spectroscopy. Zoom on the sample shows {\textbf{a)}} the situation with NIR and green light focused on an FND on glass, in dry conditions. Similarly {\textbf{b)}} shows the same but with added water while {\textbf{c)}} illustrates the FND being optically trapped in aqueous medium. {\textbf{d)}} Laser sequence for fluorescence relaxometry measurement. In green the 532 nm laser and in red the 1064 nm laser. Pulse-widths are $t_{\rm trap} = 20 \mu s$, $t_{ini} = 35 \mu s$, $t_{RO} = 15 \mu s$ while the variable waiting time $\tau$ allows to record the relaxometry at times spanning from $7.2 \mu s$ to $720 \mu s$.}
  \label{fig:setup}
\end{figure}

The dried FND sample was placed inside a mini-incubator (pecon XS 2000) in an optical setup based on a customized Thorlabs OTKB/M modular optical tweezer system. Customization included a different trapping laser, and different objective and long-distance condenser lens. Vendors and other details are provided in the Appendix, Fig.~\ref{fig:setup-details} and Table~\ref{tab:setupcomponents}. The setup is constructed around an inverted microscope configuration, equipped with white light illumination and a high-resolution camera that capture wide-field bright-field images of the sample, and two lasers for the simultaneous illumination and manipulation of the FNDs. A 
532~nm laser was utilized for the excitation of the NV centers, 
while a 1064~nm near-infrared (NIR) laser was used for FND trapping.
The two lasers were aligned and their beams combined using a dichroic mirror 
and collectively focused onto the sample through the microscope objective, as detailed in Fig.~\ref{fig:setup}, i.e., a single objective setup. By trapping an FND and then finetuning the position of the green laser to maximize the signal on the photon counter, we ensured the same focal point for both the green and NIR lasers. Additionally, acousto-optic deflectors, controlled via a Qudi~\cite{Binder2017} interface, were used to precisely control the timing and intensity of the laser beams.

As illustrated in Fig.~\ref{fig:setup}, several modes of detection were installed: i) camera detection, ii) spectrometer detection and iii) photon count detection. 
A motorized flip-mirror allows to rapidly shift between camera detection and either spectrometer or photon count detection, for recording of FND fluorescence. 
In addition, to measure the CW ODMR spectrum of FNDs, a signal generator together with an amplifier are connected to a single-turn copper loop antenna (5~mm) placed approximately 1~mm above the sample. 

\subsection{\label{sec:Methods-FNDtrapping}FND Trapping}

All trapping experiments were performed in a fluid chamber containing 50 µL of the FND aqueous suspension unless stated otherwise. The NIR laser was utilized for trapping the FNDs in the focused beam \cite{Neuman2004,Jones2015,Gennerich2017}. FND trapping was confirmed through standard analysis of histograms and power spectral density of the positions of the FND~\cite{BergSorensen2004}, with positions recorded by back focal plane interferometry on a quadrant photodiode~\cite{Allersma1998}. Examples are shown in the Supplementary information. For relaxometry, the temporal sequence of the green and trapping lasers, as illustrated in Fig.~\ref{fig:setup}d), avoids temporal overlap between the two lasers while maintaining the FND in the trap~\cite{Russell2018a}. Strictly speaking, this protocol does not constitute a conventional spin-relaxometry measurement. 
We hence will refer to the measured decay time as $T_{\rm relax}$ throughout the remainder of this paper. The green laser was pulsed with NV initialization time $t_{ini}$ of 35 µs and read-out time $t_{RO}$ of 15 µs, while the NIR laser was turned on with duration $t_{\rm trap}$ of 20 µs to trap the FNDs.  

\subsection{Sensing Experiments} 

For the experiments detecting paramagnetic species, a gadolinium solution (0.45 M in acetate) was gradually added to the FND's environment. The concentrations were incrementally adjusted, and measurements on FNDs were performed after each addition to evaluate the changes in the NV center properties.

For the pH sensing experiments, an initial 0.1 M NaOH(aq) solution at pH 13 was prepared. The pH was then progressively decreased by adding controlled volumes of a 1 M HCl solution. After each pH adjustment, the properties of the NV centers were measured while the FNDs were maintained in the optical trap with a power of 60~mW, corresponding to an intensity of order $10^7$\;W/cm$^2$.

For the temperature sensing experiments, the medium containing the FNDs was initially maintained at 37 \(^\circ\)C using an incubator. The temperature was then increased in steps of 1K, allowing the evaluation of temperature changes on the NV center properties. After each change of the set temperature of the incubator, we waited 10 minutes before the measurements were conducted.

For the sensing of paramagnetic species and pH, changes to the surrounding medium was involved. To ensure that the FND remained trapped despite disturbances in the medium, the CW NIR laser power was temporarily increased to around 100~mW while liquid was added. Subsequently, the laser power was reduced back to 60~mW for measurements with the new conditions, and the same FND.

\subsection{\label{sec:results_simuCSD} Simulations - Model of Charge State Dynamics}

\begin{figure}[ht!]
    \centering
    \includegraphics[width=1.0\linewidth]{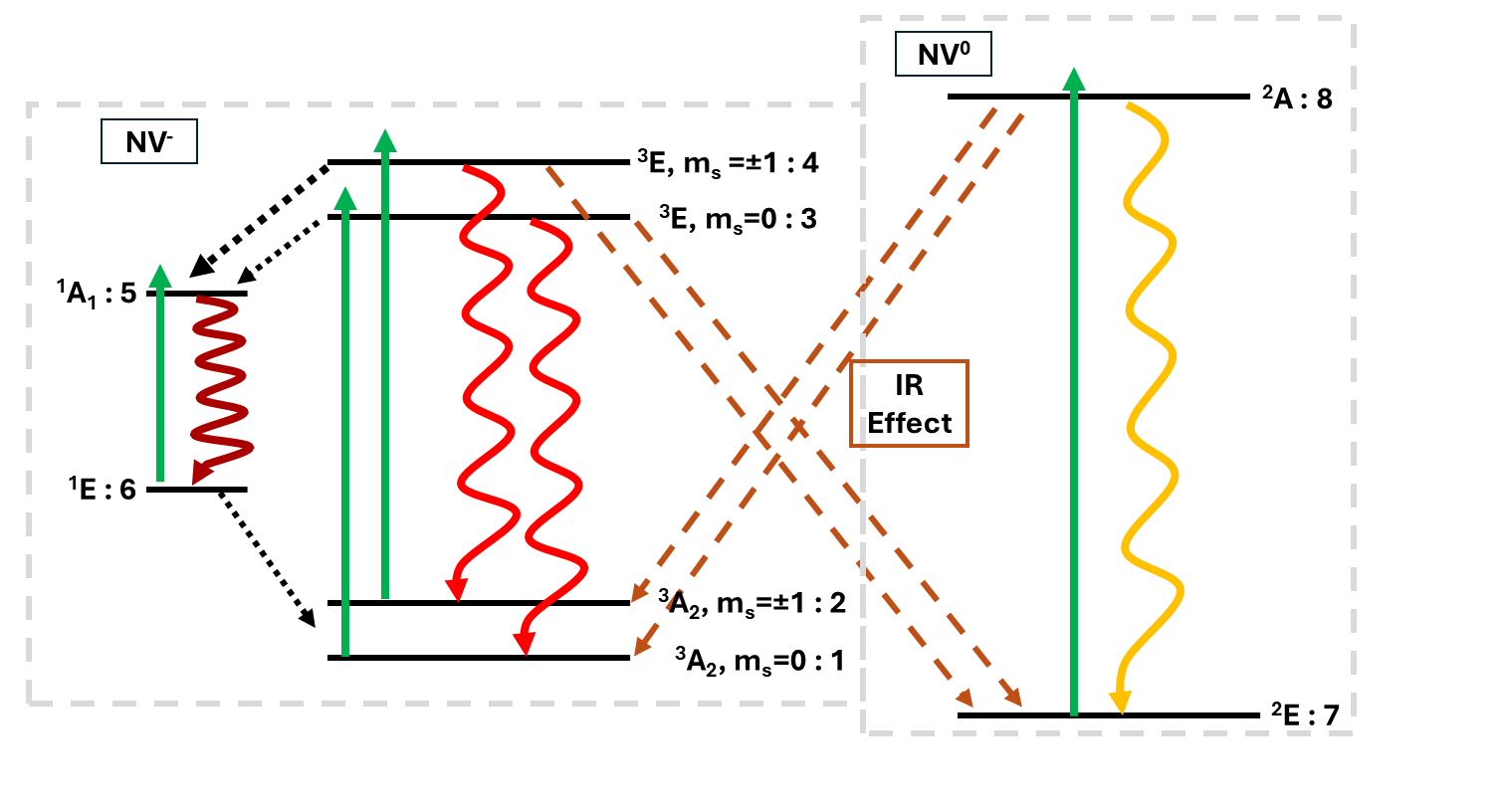}
    \caption{Energy level diagram of the NV center, showing the ground states $^3A_2, m_s=0$ (denoted as 1), $^3A_2, m_s=\pm 1$ (2) and $^2E$ (7), excited states $^3E, m_s=0$ (3), $^3E, m_s=\pm1$ (4) and $^2A$ (8), singlet shelving states $^1E$ (5) and $^1A_1$ (6), and the transitions between them. Solid arrows represent radiative transitions, dotted arrows represent non-radiative transitions, and dashed arrows represent ionization and recombination processes.}
    \label{fig:energy_levels}
\end{figure}

To understand and quantitatively model the effects of NIR laser illumination on the charge state dynamics and fluorescence properties of NV centers in FNDs, we investigate a comprehensive rate equation model \cite{Qian2022}, supplemented with insights from Ref.~\cite{Meirzada2018}. This model accounts for the various optical transitions, inter-system crossings, ionization, and recombination processes that occur when NV centers are simultaneously exposed to green and NIR laser excitation.
The model considers eight states as illustrated in the schematic energy level diagram in Fig.~\ref{fig:energy_levels}.
These states include the negatively charged NV\(^-\) and the neutrally charged NV\(^0\), with each their ground and excited electronic states, as well as inter-system crossing pathways to metastable quartet states. 
The population dynamics of the NV centers are described by a set of coupled differential equations representing the rates of change of the population probabilities $P_i$ in each state $i = 1,\ldots 8$ \cite{Qian2022}. The rates of change were adapted to our experimental situation with just two lasers, the green 532 nm and the NIR 1064 nm laser, following Ref.~\cite{Meirzada2018}. Values used in this work for the rates are provided in the appendix, Table~\ref{tab:rateequations}, and rate-equations are explicitly written in the Supplementary information. 
Simulations were carried out using the {\tt solve\_ivp} function from the {\tt sciPy} library in python.

The model assumes that populations reach a steady state within the simulation time, and that transitions happen with constant rate coefficients, independent of local temperature and other environmental factors including external magnetic or electric fields. Further, the model neglects multi-photon processes as well as higher excited states as the probabilities for these are negligible under the experimental conditions. 

\subsection{\label{sec:results_simuT} Simulation - Thermal response}
To model a potential temperature increase induced by the NIR laser on FNDs, we simulated the heat transfer. The simulation accounts for the absorption of laser energy by the FND and the subsequent heat dissipation into the surrounding medium (water or air). 
To quantify the temperature rise produced by the 1064 nm laser on a spherical fluorescent nanodiamond of radius \(R\)), we solve Fourier’s heat‑conduction equation with an internal absorption source inside the particle and pure conduction in the surrounding infinite medium (water or air):

\begin{align*}
\rho_d c_d \frac{\partial T_d}{\partial t}
      &= \frac{1}{r^{2}}\frac{\partial}{\partial r}\!\left(k_d r^{2}\frac{\partial T_d}{\partial r}\right)
         + q_{\mathrm{abs}}(r), & 0\le r < R, \nonumber\\
\rho_m c_m \frac{\partial T_m}{\partial t}
      &= \frac{1}{r^{2}}\frac{\partial}{\partial r}\!\left(k_m r^{2}\frac{\partial T_m}{\partial r}\right), & r>R,
\label{eq:heatPDE}
\end{align*}
subject to the interface and far‑field conditions
\begin{align*}
T_d(R,t)=T_m(R,t),
\end{align*}
\begin{align*}
k_d\!\left.\frac{\partial T_d}{\partial r}\right|_{r=R}
      =k_m\!\left.\frac{\partial T_m}{\partial r}\right|_{r=R}, \qquad
\lim_{r\to\infty}T_m(r,t)=T_\infty.
\end{align*}
Here \(\rho\), \(c\), and \(k\) are the density, specific heat, and thermal conductivity of the nanodiamond and the surrounding medium. The laser absorption is treated as a uniform volumetric heat source inside the FND,
\[
q_{\mathrm{abs}}(r)=
\begin{cases}
\dfrac{P_{\mathrm{abs}}}{\tfrac{4}{3}\pi R^{3}}, & r<R,\\[6pt]
0, & r\ge R,
\end{cases}
\]
where \(P_{\mathrm{abs}}=\sigma_{\mathrm{abs}}I_0\) is the absorbed optical power for a beam of intensity \(I_0\) and absorption cross‑section \(\sigma_{\mathrm{abs}}\).

Because of the small size of the FNDs, the simulation assumes a uniform temperature throughout the FND at any given time, it assumes insignificant heat loss due to thermal radiation, and constant thermal properties of the FND and the medium, including a constant convective heat transfer coefficient based on the thermal conductivity of the medium. Further, the model assumes that the medium remains in the same phase (liquid) throughout the simulation.

\section{\label{sec:Results}Results}

\begin{figure*}[tb]
    \centering
    \begin{minipage}{0.32\textwidth}
        \centering
        \includegraphics[width=\linewidth]{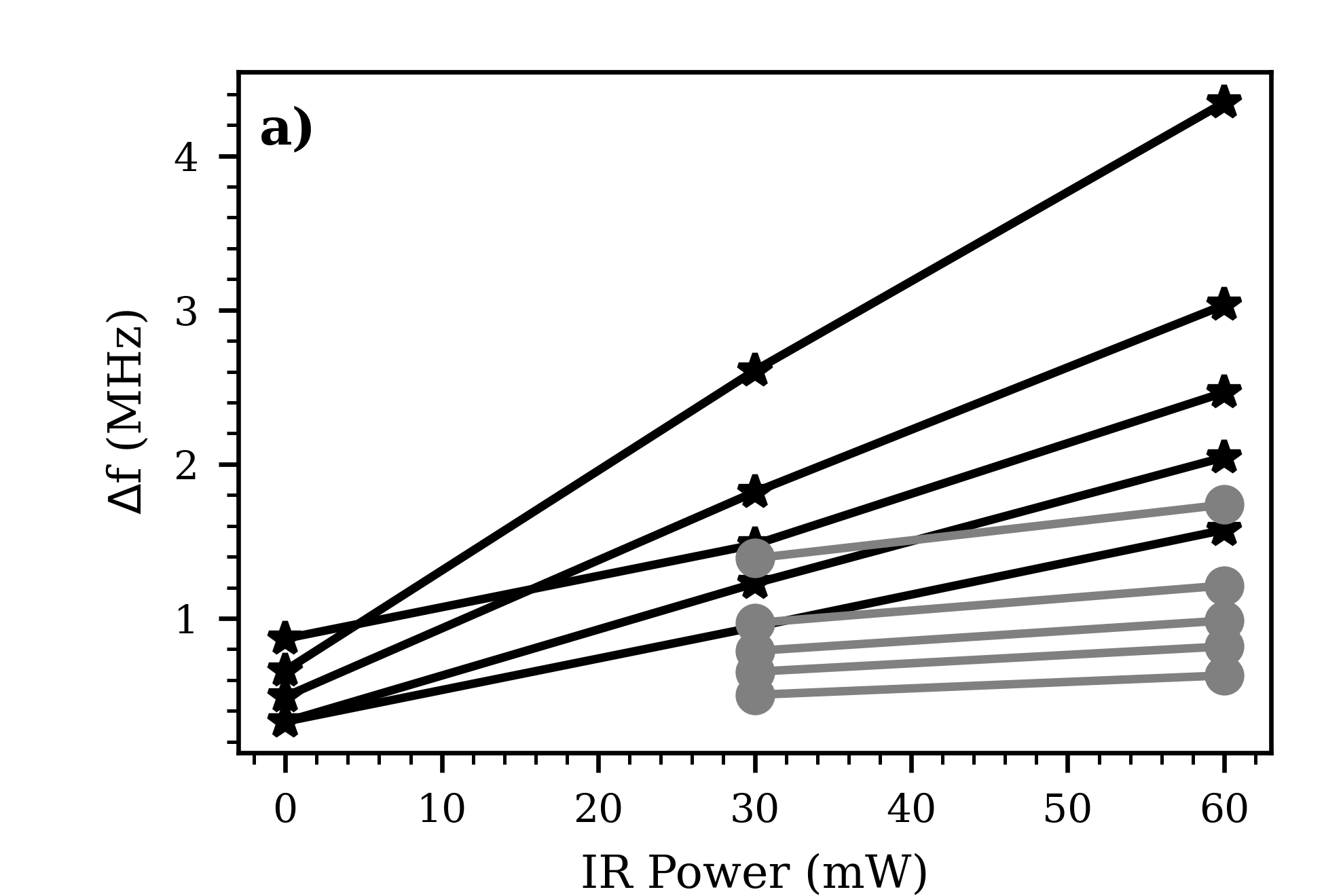}
        \label{fig:odmr1_stat}
    \end{minipage}%
    \hfill
    \begin{minipage}{0.32\textwidth}
        \centering
        \includegraphics[width=\linewidth]{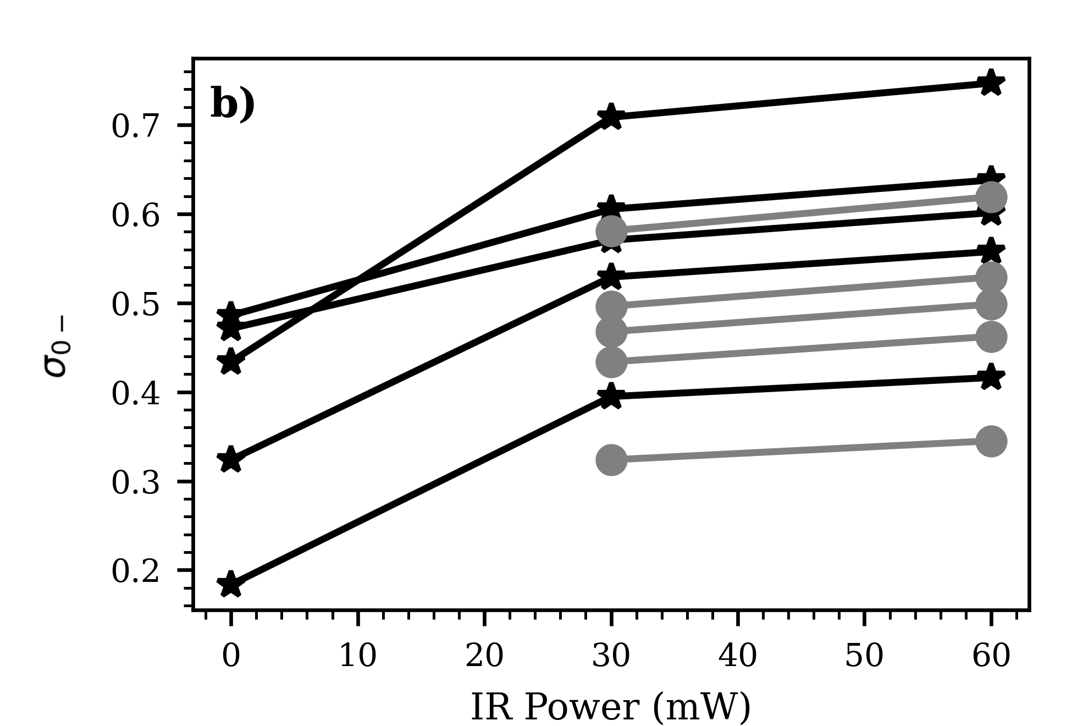}
        \label{fig:ratio_stat}
    \end{minipage}%
    \hfill
    \begin{minipage}{0.32\textwidth}
        \centering
        \includegraphics[width=\linewidth]{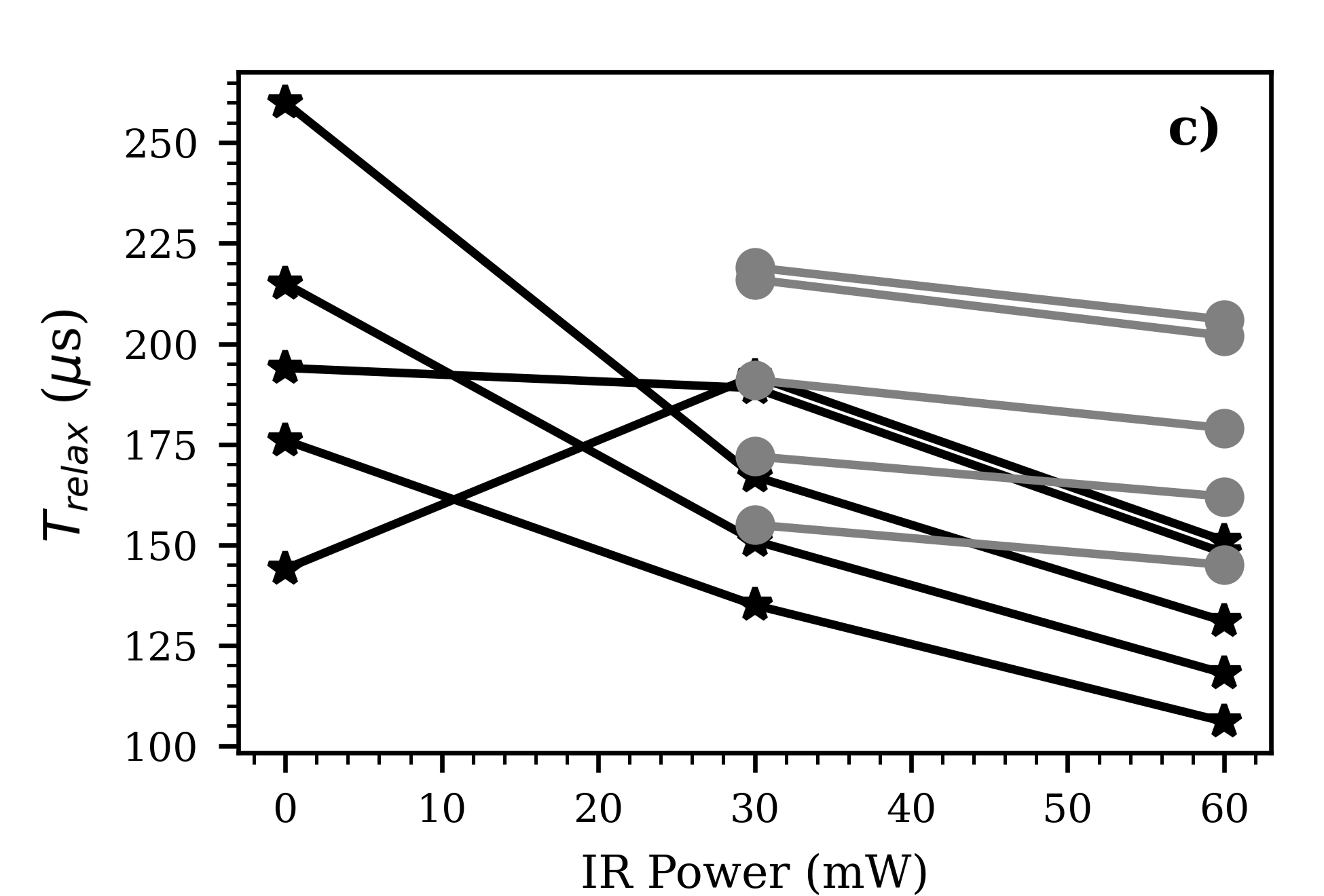}
        \label{fig:t1_stat}
    \end{minipage}
    \caption{Representative examples of variation for individual nanodiamonds.
    \textbf{(a)}~ODMR resonance frequency shift, relative to the expected resonance at 2.87~GHz,
    \textbf{(b)}~fluorescence intensity ratio $\sigma_{0-}$,
    and \textbf{(c)}~$T_{\rm relax}$ relaxometry times as functions of NIR laser power.
    Data for 5 random FNDs dry on glass (black stars) and trapped in water (gray circles) 
    with 1064\,nm NIR laser irradiation.}
    \label{fig:all_stats}
\end{figure*}

\begin{figure*}[tb]
  \centering
    \includegraphics[width=12cm]{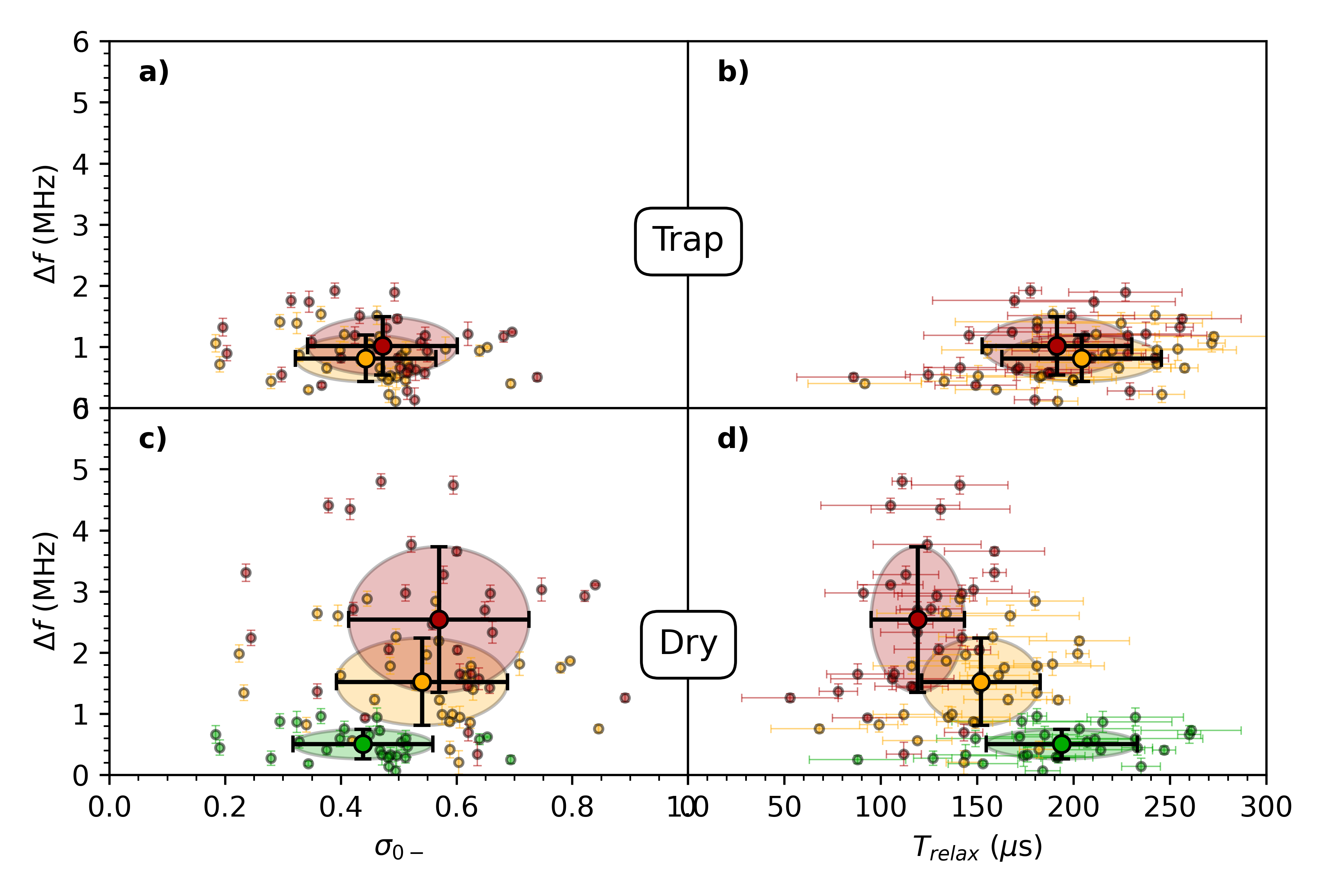}
    \caption{Summary statistics. \textbf{a)} Correlation between NIR laser power, ODMR resonance frequency shift and $\sigma_{0-}$ fluorescence intensity ratio for FNDs trapped inside water. \textbf{b)} Correlation between NIR laser power, ODMR resonance frequency shift and fluorescence relaxometry times for FNDs trapped inside water. \textbf{c)} Correlation between NIR laser power, ODMR resonance frequency shift and $\sigma_{0-}$ fluorescence intensity ratio for FNDs dry on glass. \textbf{d)} Correlation between NIR laser power, ODMR resonance frequency shift and fluorescence relaxometry times for FNDs dry on glass. Data for 30 FNDs exposed to 0~mW (green), 30~mW (orange), and 60~mW (red) of 1064~nm NIR laser irradiation. CW NIR for ODMR and spectra, pulsed NIR as in Fig~\ref{fig:setup}d) for fluorescence relaxometry. In general, we observe what appears to be a trend with differences between the three powers of NIR laser light for FNDs dry on glass, and and no or much less pronounced trend for FNDs in the optical trap in water. } 
    \label{fig:main_stat}
\end{figure*}

Our initial characterisation experiments encompass ODMR and photoluminescence (PL) spectroscopy, and fluorescence relaxometry, with and without NIR irradiation. 
PL spectrum measurement and ODMR measurement were conducted with CW NIR laser irradiation while relaxometry experiments applied the pulse protocol in Fig.~\ref{fig:setup}d).
Three conditions were investigated, either with FNDs on glass under dry conditions or with a drop of water, or FNDs in aqueous solution in the optical trap, cf.\ Fig.~\ref{fig:setup}a)-c).
The dry conditions facilitates the understanding of thermal phenomena, such as localized heating effects, and are intended as a clear reference to compare with conditions involving FNDs suspended in aqueous environments. We acknowledge that the dry condition itself lacks direct practical application in sensing contexts.
For each set of experiments, we exposed single FNDs to CW NIR laser at power levels of 0~mW, 30~mW, and 60~mW, evaluated at the position of the sample.
Fig.~\ref{fig:all_stats} show five representative examples of variation in properties from individual FNDs as NIR power increases.
As detailed below, data for FNDs on glass with a drop of water are not shown, but will be described.
Fig.~\ref{fig:main_stat} summarizes the results for 30 FNDs under dry conditions or in aqueous solution in the optical trap. Examples of the underlying measurements are provided in the Supplementary information.

To evaluate thermal effects, ODMR spectra were acquired in continuous-wave (CW) mode with constant intensities of both the 532~nm pump and 1064~nm trapping lasers and the microwave source, while sweeping the frequency of the microwave in steps of 0.3~MHz. 
Generally and for all FNDs, we observe a redshift in the resonance frequency with increasing CW NIR laser power. 
For FNDs dried on glass substrates, the resonance frequency shifted significantly to the lower frequencies. Taking 2.87~GHz as reference value for the zero field splitting, the resonance frequency shift increased in absolute value from $-0.31 \pm 0.54$~MHz at 0~mW to $-5.83 \pm 046$~MHz at 60~mW of CW NIR power. In contrast, FNDs trapped in water exhibited much smaller frequency shifts, with changes of $-0.84 \pm 0.21$~MHz at 30~mW and $-1.58 \pm 0.33$~MHz at 60~mW. 
We also observe asymmetries in ODMR contrasts corresponding to spin transitions $\Delta m_s=\pm1$, in correspondence with earlier results in literature for diamonds with ensembles of NV-centers~\cite{Mittiga2018, Broadway2019,Bauch2020}.
Guided by these insights from literature, we interpret the asymmetries to be the result of local strain heterogeneity and uneven charge distributions near the diamond surface.
 
To evaluate charge-state balance, we record the PL spectrum in the range between 550~nm to 745~nm with the spectrometer for each of the different values of the CW NIR laser power.
Upon CW NIR laser light exposure, our PL measurements revealed a significant drop in the total PL intensity.  
The curves for FNDs on glass remain quite similar when water is added, although with less effect of increase in NIR laser intensity from 30~mW to 60~mW 
and the difference between 30~mW and 60~mW is even more reduced for FNDs that are trapped in water.
To quantify the changes in relative charge state emission intensities under CW NIR laser irradiation, we determine the ratio $\sigma_{0-}$ between NV\(^0\) and NV\(^-\) emission intensities by integrating over the spectral ranges for NV\(^0\) (550--620~nm) and NV\(^-\) (655--745~nm), respectively. Compared to the analysis used in Ref.~\cite{Sow2020}, we applied a normalization factor 19/14 to account for the difference in the relative spectral ranges:

\begin{equation}
    \sigma_{0-}
    \equiv \left( \frac{\sum_{\lambda=550}^{620} I_{\lambda}}{\sum_{\lambda=655}^{745} I_{\lambda}} \right) \times \frac{19}{14}
\end{equation}
For FNDs trapped with NIR powers of 30~mW and 60~mW, we observe an averaged similar emission ratios $\sigma_{0-}$ of $0.48 \pm 0.09$. In contrast, for FNDs dried on glass, we measure an emission ratio of $0.43 \pm 0.11$ without NIR irradiation that increases to $0.57 \pm 0.15$ for a NIR laser power of 60~mW NIR. 
This is illustrated by the representative examples of fitted values for $T_{\rm relax}$ in Fig.~\ref{fig:all_stats}b) and the summary of the same for 30 individual FNDs in Fig.~\ref{fig:main_stat} parts a) and c).

Last, we conducted relaxometry experiments under different NIR laser power, to extract values for the corresponding $T_{\rm relax}$. 
For FNDs dried on glass substrates, we observed a significant decrease in the $T_{\rm relax}$ relaxation times with increasing NIR laser power. 
For FNDs trapped in water, the decrease in $T_{\rm relax}$ due to NIR irradiation was significantly less, and we hardly observe any visible change in the two relaxometry curves. 
Representative examples are shown in Fig.~\ref{fig:all_stats}c) with statistics of all results summarized in Fig.~\ref{fig:main_stat}, part b) and d).

To study the significance of the observed effects, we performed an Analysis of Variance (ANOVA) test on the $T_{\rm relax}$ relaxation times, ODMR frequency shifts, and $\sigma_{0-}$ ratios. The ANOVA results are summarized in Table~\ref{tab:anova} (one-way Kruskal–Wallis test, $\alpha = 0.05$).
\begin{table}[!ht]
   \centering
   \caption{ANOVA test results for the effect of NIR laser power on FNDs dried on glass substrates, on glass after adding water and trapped in water. For FNDs on glass, dry or with water, three groups were compared (NIR power of 0~mW, 30~mW or 60~mW) while for FNDs in the optical trap, only two groups (NIR power of 30~mW and 60~mW). The p-values indicate the significance of differences observed across NIR power levels for $T_{\rm relax}$ relaxation times, ODMR frequency shifts, and $\sigma_{0-}$ ratios.}
   \label{tab:anova}
   \begin{tabular}{|c|c|}
        \hline  
        & \textbf{p-value}\\  
        \hline\hline
        \multicolumn{2}{|c|}{\textbf{FNDs dried on glass}}\\ 
        \hline
        $T_{\rm relax}$ Relaxation Time & $< 0.0001$\\  
        \hline
        ODMR Frequency Shift & $< 0.0001$\\ 
        \hline
        $\sigma_{0-}$ Ratio & 0.0017\\
        \hline\hline
        \multicolumn{2}{|c|}{\textbf{FNDs on glass with water}}\\ 
        \hline
        $T_{\rm relax}$ Relaxation Time & $0.020$\\  
        \hline
        ODMR Frequency Shift & $0.34$\\ 
        \hline
        $\sigma_{0-}$ Ratio & $0.019$\\
        \hline\hline
        \multicolumn{2}{|c|}{\textbf{FNDs Trapped in Water}}\\
        \hline
        $T_{\rm relax}$ Relaxation Time & 0.221\\  
        \hline
        ODMR Frequency Shift & 0.073\\ 
        \hline
        $\sigma_{0-}$ Ratio & 0.371\\
        \hline
    \end{tabular}
\end{table}
For FNDs dried on glass substrates, the very low p-values (p~$< 0.01$) for all three parameters confirm that the differences observed with varying NIR laser powers are statistically significant. This indicates that NIR irradiation has a considerable impact on the NV centers' properties in air.
For FNDs on glass substrates after adding water, the 
p-values of $\simeq 0.02$ for $T_{\rm relax}$ and $\sigma_{0-}$ indicate statistical significance while the ODMR frequency shift, indicative of thermal effects, does not show statistically significant variation. 
In contrast, for FNDs trapped in water, the p-values are high (p~$> 0.07$), for all three quantities, suggesting that the differences observed across different NIR power levels are not statistically significant. 

Overall, as showed in Fig.~\ref{fig:all_stats} and Fig.~\ref{fig:main_stat}, the data suggest that the aqueous environment reduces the effect of NIR irradiation on ODMR-frequency shifts. Further, our results confirm that with specific NIR irradiation sequences and optimizing laser parameters, cf.\ Fig.~\ref{fig:setup}d) and Ref.~\cite{Russell2018a}, potential detrimental effects of NIR on $T_{\rm relax}$ relaxation can be significantly mitigated. 

\begin{figure}[ht!]
    \centering
    \includegraphics[width=9cm]{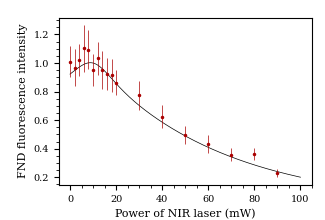}
    \caption{\textbf{Total fluorescence intensity from NV centers as a function of NIR laser power.} Black: Theory, Red: Experimental observations with errorbars, obtained as a statistical average over results from 10 different FNDs with $\sigma_{0-}$ within the range of $\sigma_{0-} = 0.42 \pm 0.04$.}
    \label{fig:fluorescence}
\end{figure}

The rate equation simulation reveal that at low CW NIR laser powers, the population of NV\(^-\) centers increases slightly, while the NV\(^0\) population decreases. This is attributed to enhanced recombination processes facilitated by the NIR photons, which convert NV\(^0\) centers back to NV\(^-\). Specifically, up to a CW NIR power of approximately 10~mW, the NV\(^-\) population increases by about 8\% compared to the case without CW NIR illumination.
However, as the CW NIR power increases beyond this threshold, the trend reverses. The NV\(^-\) population begins to decrease, and the NV\(^0\) population increases. At a CW NIR power of 60~mW, the NV\(^-\) population decreases by approximately 10\% from its maximum value, while the NV\(^0\) population increases correspondingly. This inversion is due to the saturation of recombination processes and the dominance of ionization processes at higher CW NIR powers, which ionize NV\(^-\) centers back to NV\(^0\). 
Consequently, the total fluorescence intensity from the NV centers first increases then decreases with increasing CW NIR power (Fig.~\ref{fig:fluorescence}). This decrease is primarily due to the reduced population of NV\(^-\) centers and the lower fluorescence quantum yield of NV\(^0\) centers. The fluorescence from NV\(^-\) centers dominates the emission under low CW NIR power, but as the CW NIR power increases, the contribution from NV\(^0\) centers becomes more significant. 
At low CW NIR powers, the $\sigma_{0-}$ ratio is low due to the predominance of NV\(^-\) fluorescence. As the CW NIR power increases, the ratio rises slowly, indicating a relative increase in NV\(^0\) emission. This change in the fluorescence ratio affects the overall spectral characteristics and can influence the detection sensitivity in applications relying on NV\(^-\) fluorescence.
The trends observed in the simulation align with the experimental results. The initial increase and subsequent decrease of the NV\(^-\) population with increasing NIR power explain the non-monotonic behavior of fluorescence intensity (Fig.~\ref{fig:fluorescence}) and $\sigma_{0-}$ ratio (Fig.~\ref{fig:main_stat}a,c)) observed experimentally. 

The simulation of thermal effects considered the temperature change of an FND during a 500~nanosecond NIR laser pulse followed by a cooling period after the laser is turned off. The results show that the FND's temperature increases during the laser pulse and decreases afterward, illustrating the transient thermal response to short laser excitation. In our experimental pulse sequence, the NIR pulse was longer (20$\mu$s), in the simulation the choice of 500~ns has been done to show the temperature increase before saturation; we observe saturation after 75~ns for the FND in air and 435~ns for the FND in water.
The simulations show that in water, the temperature increases slightly and steadily during the laser pulse and rapidly returns to ambient temperature within nanoseconds after the laser is turned off.
Contrary, in air the temperature increase is more pronounced, it is roughly 20 times larger than that in water, and the temperature 
decreases more slowly than in water after the laser is turned off.

The results of the simulation of thermal effects align with experimental observations where NIR illumination causes notable thermal effects in air and much less in aqueous environments, cf.\ Fig.~\ref{fig:all_stats}-Fig.~\ref{fig:main_stat}. 
A direct comparison of Fig.~\ref{fig:main_stat}c)--d) with Fig.~\ref{fig:main_stat}a)--b) shows that the environment (air versus water) is the controlling parameter when it comes to the thermal response.  
In air, the ODMR resonance shifts by $-5.8\pm0.5$\,MHz and $T_{\rm relax}$ falls by $(44\pm6)\%$ when the CW NIR power is raised from~0 to~60\,mW, whereas in water the corresponding changes are only $-1.6\pm0.3$\,MHz and $(2\pm5)\%$, respectively.  
By contrast, the NV$^{0}$/NV$^{-}$ fluorescence intensity ratio $\sigma_{0-}$ responds similarly to NIR power in both media (Fig.~\ref{fig:main_stat}a,c; projection on the ordinate axis), confirming that charge-state kinetics are governed by photon flux rather than heat exchange.  

For completeness we note that NV$^{-}$ centres possess a weak
intersystem transition between the metastable singlet states
$^1E\,(p_6)\!\rightarrow\!{}^1A_{1}\,(p_5)$ with a zero-phonon line at
$\lambda_{\mathrm{ZPL}}=1042$ nm and a room-temperature homogeneous
width of only a few\,GHz
($\sim10^{-4}$\,nm)\,\cite{Rogers2015}.
Our trapping beam at 1064 nm is therefore red-detuned by  
$\Delta\lambda = 22$ nm, corresponding to 
$\Delta\nu \simeq 6$ THz ($\simeq$ \,25 meV), i.e.\ \(\sim10^{3}\) times the natural linewidth and well outside the vibrational sideband of the singlet manifold.
The effects reported in this work (PL quenching, ODMR shifts, $T_{\rm relax}$
variations) therefore most likely arise from photothermal heating and photo-ionisation dynamics, not from resonant excitation of the
1042 nm line.

\subsection{\label{sec:BIOseusning}Sensing Capabilities of Trapped FNDs}

\begin{figure*}[tb]
  \centering
    \includegraphics[width=12cm]{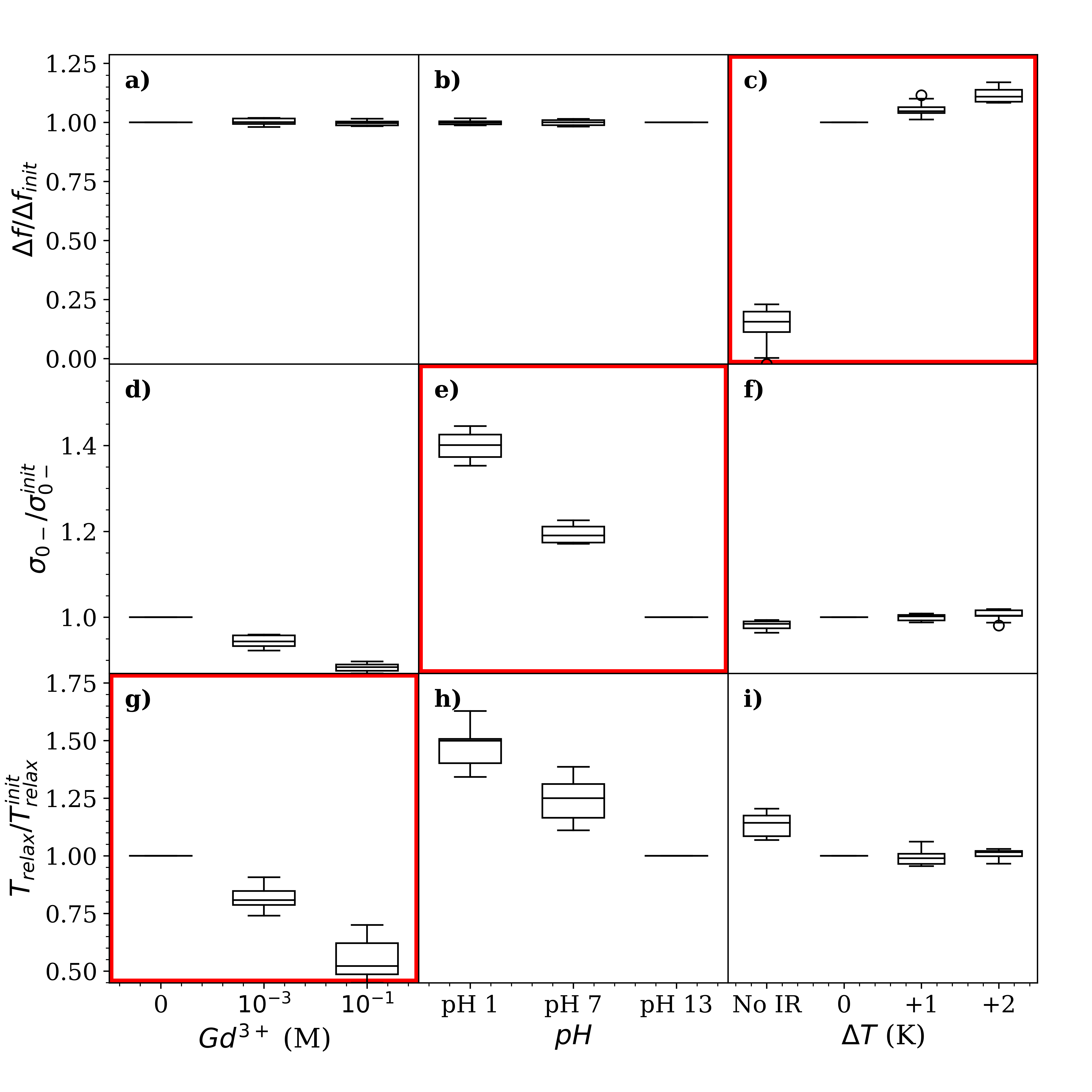}
    \caption{Variation in three sensing parameters when subject to controlled variation of paramagnetic species, pH, and temperature. a) ODMR resonance frequency shift VS Gd\(^{3+}\) concentration. b) ODMR resonance frequency shift VS pH value. c)  ODMR resonance frequency shift (relative to the expected 2.87\;GHz) VS Temperature increase (relative to 37\;$^\circ$C). d) Fluorescence relaxometry time $T_{\rm relax}$ VS Gd\(^{3+}\) concentration. e) $T_{\rm relax}$ VS pH value. f)  $T_{\rm relax}$ VS Temperature increase. g) $\sigma_{0-}$ fluorescence intensity ratio VS Gd\(^{3+}\) concentration. h) $\sigma_{0-}$ fluorescence intensity ratio VS pH value. i)  $\sigma_{0-}$ fluorescence intensity ratio VS Temperature increase. The boxplot represent results for 10 FNDs that were trapped for each condition with 60 mW 1064 nm CW NIR laser. In each case, results for 0\;M Gd\(^{3+}\), pH 13 and 0\;K trapped were used as reference values for the individual FNDs. The red squares emphasize the optimal measurement method for each sensing parameter. 
    }
    \label{fig:bio_stat}
\end{figure*}

In addition to the multi-modal characterization presented above, the potential of optically trapped FNDs with NV centers for sensing applications were addressed. Experiments focused on three key parameters: Gd\(^{3+}\) concentration, pH, and temperature, and the same three measurement modalities (PL spectroscopy, ODMR, and optical relaxometry) as before. 
Results are summarized in Fig.~\ref{fig:bio_stat}.

The temperature response of nanodiamonds with NV centers was evaluated under three conditions: FNDs on glass substrates with no NIR laser irradiation, trapped FNDs with no temperature increase, and trapped FNDs with temperature increases of +1~K and +2~K. As expected, a temperature-dependent shift in the ODMR resonance frequency was observed, with lower frequencies at higher temperatures (Fig.~\ref{fig:bio_stat}c)). This shift is attributed to the well-known temperature dependence of the NV center's zero-field splitting parameter \(D\), which decreases at a rate of approximately \(-74~\text{kHz/K}\) \cite{Acosta2010}. On the contrary, no significant changes were detected in $T_{\rm relax}$ values, indicating that spin-lattice relaxation is relatively insensitive to temperature variations within the tested range (Fig.~\ref{fig:bio_stat}i)). Further, no significant changes in $\sigma_{0-}$ were observed with temperature, suggesting that charge state dynamics are minimally affected by temperature within the range investigated here (Fig.~\ref{fig:bio_stat}f)).
ODMR's temperature sensitivity highlights its utility for high-resolution nanoscale thermometry. The absence of significant $T_{\rm relax}$ or optical spectral changes simplifies calibration, making ODMR a standalone method for temperature measurements.

The impact of environmental pH on the properties of NV centers was evaluated using buffer solutions with pH values of 1, 7, and 13.
\begin{figure}[tb]
  \centering
    \includegraphics[width=9cm]{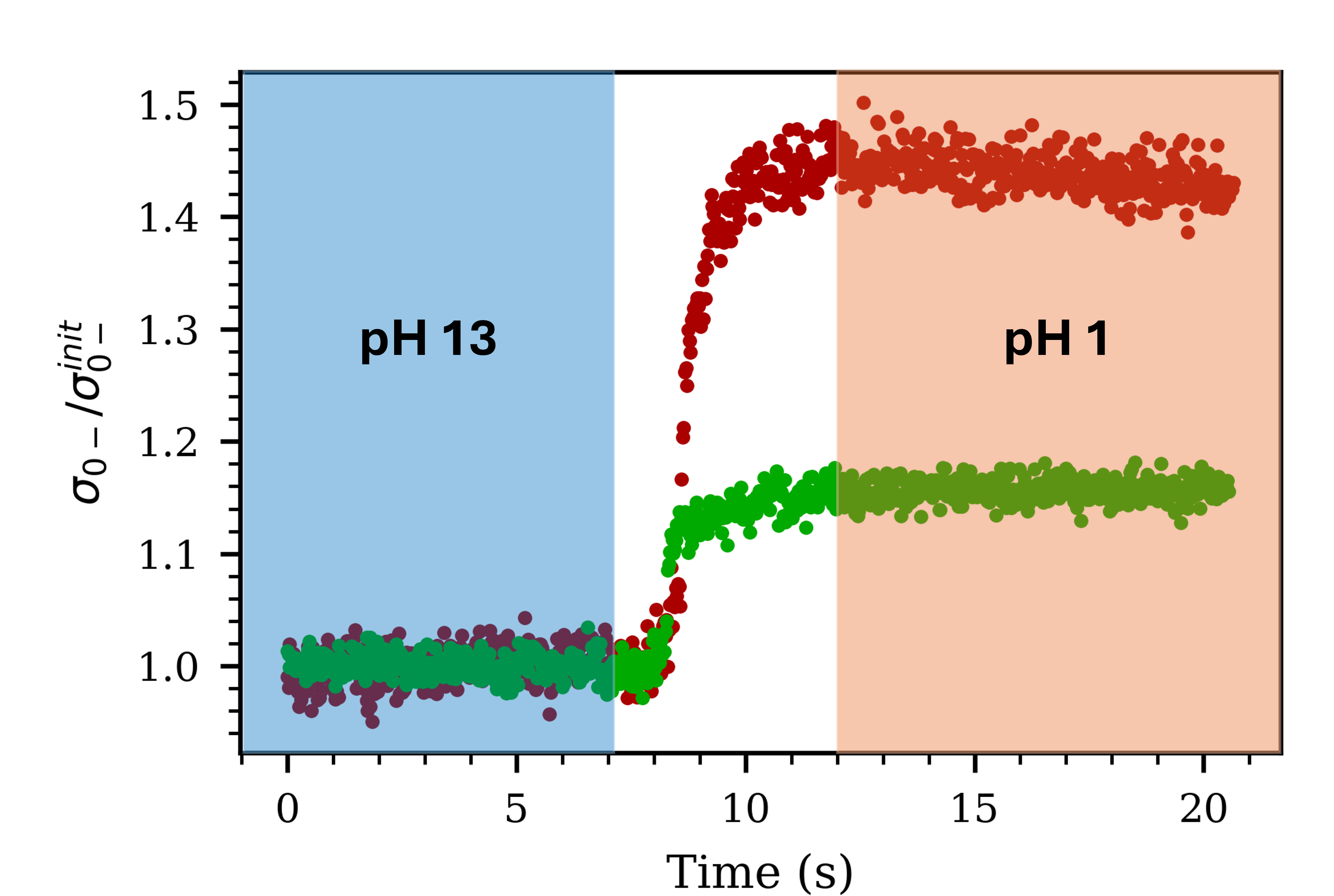}
    \caption{\textbf{$\sigma_{0-}$ fluorescence intensity ratio during pH value changes.} Measurement of the same FND while changing from pH13 to pH1 by adding HCl solution. Red: FND trapped with 60 mW 1064 nm CW NIR laser, Green: FND on glass without NIR laser.}
    \label{fig:ph_ratio}
\end{figure}
No significant changes in ODMR resonance frequency (Fig.~\ref{fig:bio_stat}b)) were detected across the tested pH range, indicating minimal impact of pH on electronic spin states. A decrease in $T_{\rm relax}$ values was observed as pH increased, indicating a dependence on surface protonation dynamics (Fig.~\ref{fig:bio_stat}h)). This is consistent with reports that high pH conditions stabilize spin relaxation dynamics. A decrease in the $\sigma_{0-}$ ratio with increasing pH (Fig.~\ref{fig:bio_stat}e)) and Fig.~\ref{fig:ph_ratio}) suggests that alkaline conditions favor the NV\(^-\) charge state. It may be attributed to the deprotonation of surface functional groups such as carboxyl (–COOH) and hydroxyl (–OH) groups present on the nanodiamond surface. Deprotonation of these groups under alkaline conditions generates negatively charged surface sites, thereby enhancing electron donation to nearby NV centers and favoring the stabilization of the NV$^-$ charge state \cite{karaveli2016}. The difference in ratio change between FNDs dry on glass without an NIR laser and FNDs trapped with a 60 mW CW IR laser (Fig.~\ref{fig:ph_ratio}) shows a clear advantage of using FNDs in suspension for sensing. 
Another advantage of the optical trap is that effects of individual variation between FNDs from a batch is reduced since measurements can be made relative to an initial measurement with the same FND.

To further explore the dependence of the NV charge state on pH, Fig.~\ref{fig:ph_ratio} presents the evolution of the $\sigma_{0-}$ ratio when varying pH from 13 to 1 for the same FND, both under trapping conditions and without NIR laser exposure. The trapped FNDs in suspension exhibit a significantly stronger modulation of $\sigma_{0-}$ compared to the same particle measured on glass. This enhanced responsiveness highlights the improved surface accessibility and chemical exchange in solution, reinforcing the utility of optically trapped FNDs for pH nanosensing.  
Overall, the results confirm that  $\sigma_{0-}$ serves as a reliable and sensitive optical proxy for probing local pH variations in biological or environmental settings.

The sensitivity of NV centers to paramagnetic gadolinium ions (Gd\(^{3+}\)) was investigated through measurements at 3 different values of the concentration.
\begin{figure}[tb]
  \centering
     \includegraphics[width=9cm]{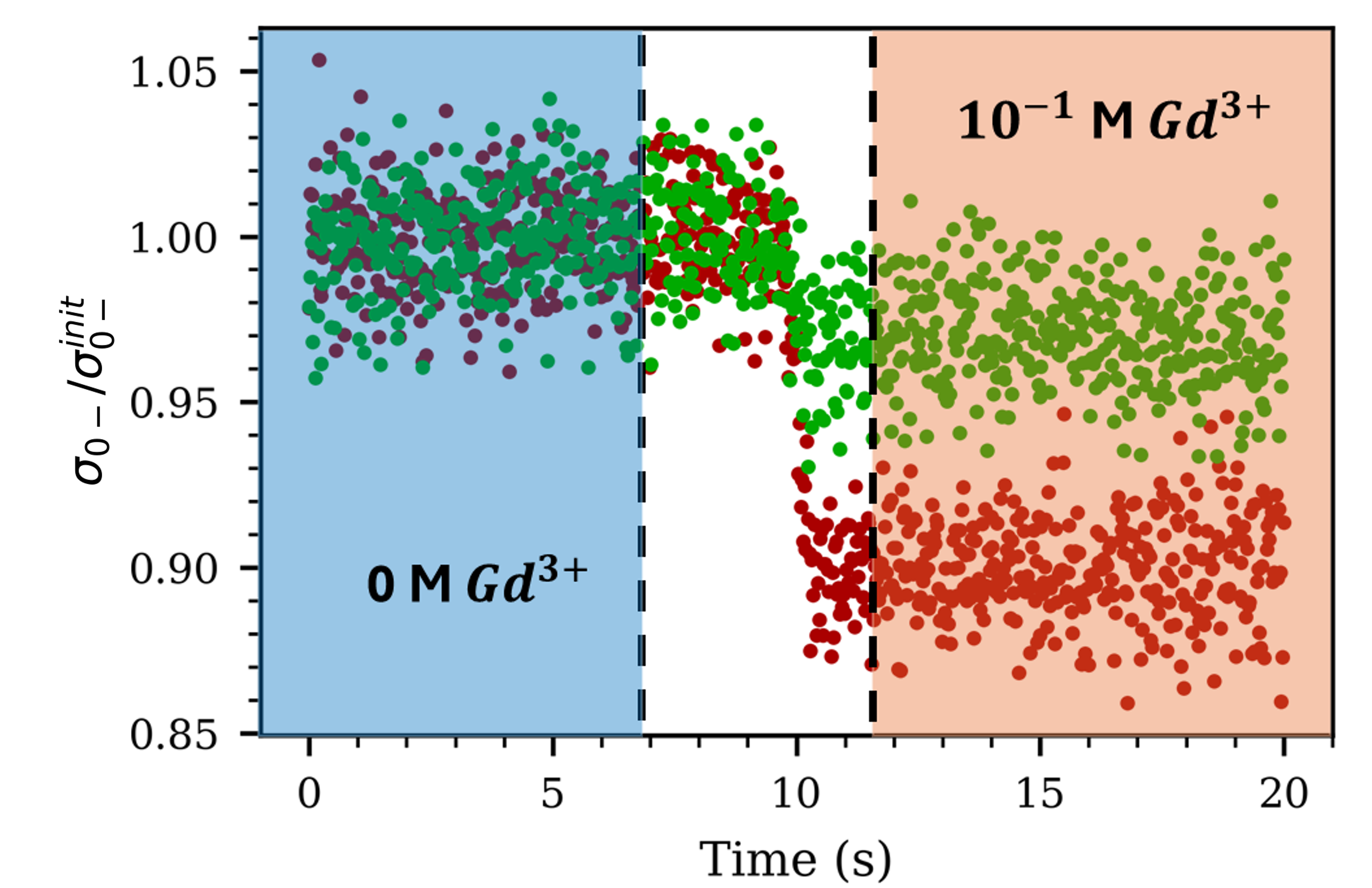}  
    \caption{\textbf{$\sigma_{0-}$ fluorescence intensity ratio during gadolinium concentration changes.} Measurement of the same FND while changing gadolinium concentration from 0M to 0.1M  by adding a gadolinium solution (0.45 M in acetate). Red: FND trapped with 60 mW 1064 nm CW NIR laser, Green: FND on glass without NIR laser.}
    \label{fig:gd_ratio}
\end{figure}
No significant shifts or changes in ODMR resonance frequency were detected, suggesting limited interaction between Gd\(^{3+}\) ions and the electronic spin states of the NV centers under the tested conditions (Fig.~\ref{fig:bio_stat}a)). A decrease in $\sigma_{0-}$ was observed (Fig.~\ref{fig:bio_stat}d) and Fig.~\ref{fig:gd_ratio}). This can be attributed to charge state dynamics, where Gd\(^{3+}\) ions interact with the FND surface, altering the local electrochemical potential and promoting ionization of NV\(^-\) to NV\(^0\) \cite{Rondin2010}.  The difference in ratio change between FNDs dry on glass without an NIR laser and FNDs trapped with a 60 mW CW IR laser (fig.~\ref{fig:gd_ratio}) shows a clear advantage of using FNDs in suspension for sensing. Finally, a significant decrease in $T_{\rm relax}$ values was observed with increasing Gd\(^{3+}\) concentration (Fig.~\ref{fig:bio_stat}g)). This result aligns with the quenching effect of paramagnetic ions, which enhance spin-lattice relaxation by introducing magnetic noise to the local environment \cite{Tetienne2013,Rendler2017,Martinez2020}.
 
These findings demonstrate the potential of optically trapped nanodiamonds with NV centers for detecting paramagnetic species like Gd\(^{3+}\). 
The fluorescence relaxometry provides a robust method for quantifying Gd\(^{3+}\) in biological and environmental contexts. However, the effect on charge state dynamics also imply that conclusions from fluorescence relaxometry in otherwise uncharacterized environments should be made with caution, and at best complemented with fluorescence spectrometry. For completeness, we observe that by adding Gd\(^{3+}\) to a concentration of 0.1~M, the pH value in the solution increases slightly (by of order 1-2~pH units) which is consistent with our previous observations on the effect of pH on the fluorescence of FNDs.

\section{\label{sec:discuss}Discussion}

The shift in the ODMR resonance frequency with increasing NIR power, as shown in 
Fig.~\ref{fig:all_stats}a), and Fig.~\ref{fig:main_stat}, suggests an increase in temperature within the FNDs. This is consistent with the known temperature dependence of the zero-field splitting parameter $D$ of the NV center, which decreases with increasing temperature at a rate of approximately $-74$~kHz/K~\cite{Acosta2010, Toyli2012} and corresponds with previous suggestions of the heating effect of NIR light~\cite{Lai2013}. However, the extent of the thermal effect is significantly influenced by the surrounding medium's properties. Media with higher thermal conductivity and heat capacity, such as water, dissipate heat more efficiently than air, resulting in smaller temperature increases in the FNDs. This is supported by our temperature simulations, which show much lower temperature rise in water compared to air during NIR illumination. Consequently, FNDs trapped in water exhibit much smaller ODMR frequency shifts with increasing NIR power.

In addition to thermal effects, the NIR laser alters the charge-state dynamics of the NV centers. Our simulations of charge-state dynamics indicate that at low NIR power levels, there is a slight increase in the NV\(^-\) population due to enhanced recombination processes facilitated by NIR photons. However, as the NIR power increases beyond a threshold (approximately 10~mW with our parameters), the NV\(^-\) population begins to decrease while the NV\(^0\) population increases. This inversion is attributed to the saturation of recombination processes and increased ionization rates at higher NIR powers.
As a result, 
the total fluorescence intensity decreases with increasing NIR power, as shown in Fig.~\ref{fig:fluorescence}. This decrease can be explained by the reduced NV\(^-\) population and the lower fluorescence quantum yield of NV\(^0\) centers~\cite{Chapman2010}. The increase in $\sigma_{0-}$ with higher NIR power, depicted in Fig.~\ref{fig:all_stats}b) and Fig.~\ref{fig:main_stat}, further confirms the shift in charge state distribution. Additionally, the increased NV\(^0\) background fluorescence dilutes the ODMR contrast since NV\(^0\) centers do not exhibit spin-dependent fluorescence variations. This is reflected in the decreased ODMR contrast observed in our measurements (data not shown).
Charge-state fluctuations also introduce additional pathways for spin relaxation, contributing to the reduction in $T_{\rm relax}$ times~\cite{Aslam2013}.
With an all-optical protocol for relaxometry as used here, the results for $T_{\rm relax}$ presented inherently involves potential inaccuracies due to photoionization induced by the green laser itself, thus affecting the NV\(^{-}\) charge states. To address this limitation, measurement protocols incorporating microwave pulses explicitly designed to disentangle spin dynamics from photoionization effects have been developed. Examples include the microwave $\pi$-pulse method introduced by Ref.~\cite{Jarmola2023}, as well as alternative sequences described in Ref.~\cite{Cardoso2023}. 

Our results indicate that although both thermal effects and charge-state dynamics significantly impact the properties of the NV centers under NIR illumination, their relative contributions depend on the environmental conditions. In FNDs dried on glass substrates, the pronounced ODMR frequency shifts suggest that thermal effects dominate due to inefficient heat dissipation in air. In contrast, for FNDs trapped in water, the minimal temperature increase reduces thermal effects, making charge-state dynamics more prominent in influencing the NV centers' behavior. The increase in the $\sigma_{0-}$ ratio and the decrease of fluorescence relaxometry times with NIR power without significant ODMR frequency shifts (Fig.~\ref{fig:main_stat}) indicate that charge-state transitions are the primary mechanism affecting the fluorescence and spin properties in aqueous environments.

While thermal effects and charge-state dynamics provide a comprehensive explanation for our observations, alternative mechanisms should be considered. Multi-photon absorption processes could directly excite the NV centers without significant heating, yet, the relationship between ODMR frequency shift and NIR power (Fig.~\ref{fig:main_stat}) is contrary to this explanation.
Moreover, the significant differences in the effects of NIR irradiation between FNDs in air and water emphasize the critical role of the surrounding medium. The statistical analysis using ANOVA (Table~\ref{tab:anova}) confirms that the differences observed in FNDs dried on glass are statistically significant, while those in FNDs trapped in water are not.
It should also be emphasized that while our results are consistent with an interpretation of ODMR frequency shift being proportional to a temperature shift, the interpretation may be more complex since changes in local electrical field also implies shifts of the ODMR frequency~\cite{Dolde2011}, an effect that we cannot completely exclude in our experiments. Similarly,  literature suggests NIR-induced dark state transitions with strong NIR pulses~\cite{Geiselmann2013,Cao2025a,Cao2025b}, and including these extra states in our theoretical model could provide more detailed understanding of our results.

In our opinion, the data presented in this work establish a clear hierarchy of influences:
(i) the thermal conductivity and heat capacity of the surrounding medium dictate the magnitude of the NIR-induced temperature rise and therefore the thermally induced shifts in ODMR frequency and the shortening of $T_{\rm relax}$;  
(ii) the NIR photon flux sets the rate of photo-ionisation and recombination, which in turn controls the NV$^{0}$/NV$^{-}$ balance and the fluorescence contrast.  
Optical trapping itself plays a mainly mechanical role--holding a single FND at a fixed position—while 
potential effects on spin and charge-state dynamics depend on the wavelength of the trapping laser. 

This distinction is important for two reasons.  
First, it tells experimentalists which knob to turn: to suppress thermal artefacts one changes the medium (e.g.\ buffer composition, flow, heat sinking) rather than the trap geometry and wavelength.  
Second, it generalises our findings to intracellular sensing, where the cytoplasm’s thermal properties resemble those of water: the modest $\lvert\Delta f\rvert<2$ MHz and unchanged $T_{\rm relax}$ observed {\em in vitro} set an upper bound on in-cell NIR heating, validating the use of trapped FNDs for quantitative biosensing. 

The results presented in the sensing section confirm that optical trapping of FNDs provide a valuable add-on that allows to enhance the application potential of biosensing with NV-centers in nanodiamonds. In particular, Fig.~\ref{fig:bio_stat}a-c) demonstrates that the frequency shift in ODMR is a good stand-alone measure for a temperature shift also for optically trapped FNDs, while parts f) and i) illustrate that changes in the surrounding temperature within a range that is relevant for biological systems does not appear to alter the readings provided by the fluorescence intensity ratio $\sigma_{0-}$ or the relaxation time, $T_{\rm relax}$. 

\section{\label{sec:conclusion}Conclusion}

This work investigates the effects of near-infrared (NIR) laser irradiation on fluorescent nanodiamonds (FNDs) containing nitrogen-vacancy (NV) centers, offering a comprehensive overview of the interplay between thermal effects, charge-state dynamics, and their combined influence on sensing capabilities. By employing an integrated experimental approach combining photoluminescence spectroscopy, optically detected magnetic resonance (ODMR), and fluorescence relaxometry, this study elucidates how varying NIR laser powers impact the fluorescence, spin properties, and stability of NV centers in diverse environmental conditions.

A key finding is the dual role of NIR irradiation: in dry environments, thermal effects dominate, resulting in significant shifts in ODMR frequencies and reductions in relaxation times due to localized heating. In contrast, for FNDs that are fully submerged in aqueous environments, the surrounding water appears to mitigate these thermal effects through efficient heat dissipation, allowing charge state dynamics to take precedence. The increased $\sigma_{0-}$ ratio and its implications for sensing applications underscore the critical role of the surrounding medium in tailoring NV center behavior. The study also demonstrates the sensitivity of optically trapped FNDs to pH variations, temperature fluctuations, and paramagnetic species, establishing these nanodiamonds as versatile tools for nanoscale sensing.

The findings indicate that optimizing NIR laser parameters, including power levels and pulsing sequences, can significantly enhance sensor performance while minimizing adverse effects such as fluorescence quenching and spin relaxation. Additionally, the demonstrated ability to detect and differentiate environmental factors such as pH and gadolinium ions highlights the potential of NV centers for applications in biological sensing and environmental monitoring.

However, this study also reveals critical uncertainties and challenges that must be addressed to fully exploit the potential of NV-based FNDs. Long-term stability under extended NIR irradiation remains an open question, particularly regarding potential degradation of fluorescence and spin properties. The role of surface functionalization in enhancing NV center stability and sensing accuracy requires further exploration, as does the potential for alternative NIR wavelengths or advanced laser modulation techniques to improve performance. The effects of complex biological environments, including interactions with proteins, ions, and other cellular components, are yet to be fully understood.

Theoretical modeling, while providing valuable insights into charge state dynamics and thermal effects, must be expanded to include the complexities of real-world applications. Future work could explore the integration of complementary sensing techniques, such as fluorescence lifetime imaging or advanced magnetic resonance methods, to provide a a real-time sensing protocol inside cells by using optically trap FNDs in the cellular medium. 

\section*{Acknowledgments}
This work was supported by the Independent Research Fund Denmark (grant no.\ 0135-00142B) and the Novo Nordisk Foundation (grant no.\ NNF20OC0061673). We acknowledge discussions with Maabur Sow and Fedor Jelezko, supported by ERC through HyperQ Project (SyG 856432). We further acknowledge insights provided through the student project work of Kristian Lambertsen and Nicolai Kongstad, as well as Suyash Amzare, including an original sketch of the experimental setup providing the basis for Fig.~\ref{fig:setup-details}. 

\bibliographystyle{plainnat}
\bibliography{export_new}


\vspace*{8mm}

\section*{Appendix}

\begin{figure*}
\centering
  \includegraphics[width=0.8\textwidth]{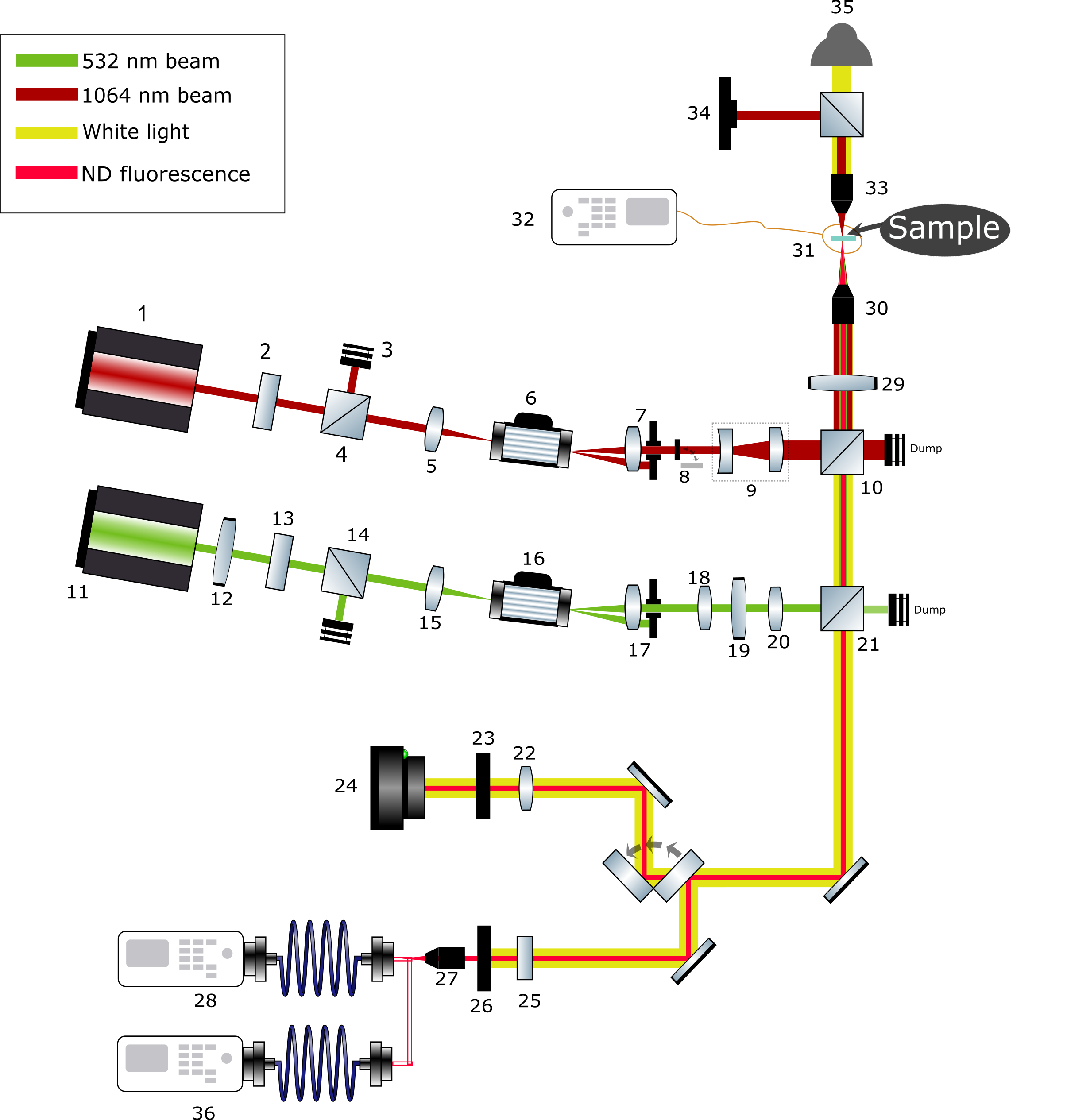}
  \caption{Detailed illustration of the optical setup with the most important lasers, optical lenses, filters and equipment. The components are specified in \protect{Table~\ref{tab:setupcomponents}}.}
  \label{fig:setup-details}
\end{figure*}

We specify in Table~\ref{tab:setupcomponents} all part numbers in the optical setup, Fig.~\ref{fig:setup-details}.

\begin{table*}
\centering
\caption{Components of the optical setup, Fig.~\ref{fig:setup-details}. Filters at position 23 and 26 are switchable. \label{tab:setupcomponents}}
\begin{tabular}{l l l}
\hline
\textbf{Ref. Number} & \textbf{Component} & \textbf{Model} \\
\hline
1 & IR laser & Cobolt - Rumba 3000 \\
\hline
2 & IR polarizer & ThorLabs Multi-order polarizer \\
\hline
3 & IR beam dump & ThorLabs BTC30 \\
\hline
4 & IR PBS & ThorLabs CCM1-PBS253/M \\
\hline
5 & Lens & ThorLabs LA1708-B-ML \\
\hline
6 & Acousto-Optic Deflector & AA Optoelectronics MT200  - A0.5 - 1064nm \\
\hline
7 & Iris Aperture & ThorLabs ID12Z/M \\
\hline
9 & Beam expander & Part of Thorlabs OTKB/M\\
\hline
10 & Dicroic mirror & DMSP805R (in ThorLabs DFM1/M) \\
\hline
11 & Green laser & Verdi G-Series - Coherent \\
\hline
12 & Optical isolator & EOT 04-532-00113 \\
\hline
13 & Green polarizer & ThorLabs Multi-order polarizer \\
\hline
14 & Green PBS & ThorLabs VA5-532/M \\
\hline
15 & Lens & ThorLabs LA1131 - A - ML \\
\hline
16 & Acousto-Optic Deflector & Brimrose TEM-110-25-530\\
\hline
17 & Iris Aperture & ThorLabs SM1D12D \\
\hline
18 & Lens & ThorLabs LA1131 - A - M \\
\hline
19 & Iris Aperture & ThorLabs P30D \\
\hline
20 & Lens & ThorLabs LA1509 - A \\
\hline
21 & Dicroic mirror & Semrock NFD01-532-25x36 (in Thorlabs DFM1/M)  \\
\hline
22 & Lens & ThorLabs LA1484 - A \\
\hline
23 & Filters & Brightline Basic 460/60 - ThorLabs MF630-69 Emission filter \\
\hline
24 & Camera & Teledyne Prime 95B \\
\hline
25 & Filter & ThorLabs NE20A \\
\hline
26 & Filters & Brightline Basic FF01-630/52 - ThorLabs FELH 0650 \\
\hline
27 & Objective lens & Olympus PLN 10X \\
\hline
28 & Single Photo counting module & Excelitas SPCM-AQRM-13-FC \\
\hline
29 & Iris aperture & ThorLabs SM1D12D \\
\hline
30 & Objective lens & Nikon Plan Apo 100X \\
\hline
31 & Microwave Antenna & Copper wire \\
\hline
32 & Microwave Generator & Agilent N5181A \\
\hline
33 & Condensor & Mitutuyo Plan Apo 100X \\
\hline
34 & Quadrant photoelectric detector & ThorLabs PDQ80A \\
\hline
35 & White lamp & Part of Thorlabs OTKB/M\\
\hline
36 & Spectrometer & Ocean Optics QE Pro \\
\hline
\end{tabular}
\end{table*}

In literature, a varying range of values for the rate constants used in the numerical model corresponding to Fig.~\ref{fig:energy_levels} exist. We applied the relevant values from Ref.~\cite{Qian2022} and combined them with parameters from~\cite{Meirzada2018} for the four ionization and recombination rates as these authors use the same laser wavelengths as us in their protocol. Below, in Table~\ref{tab:rateequations}, we provide the values used in the simulations of Section~\ref{sec:results_simuCSD}.

\begin{table*}
\centering
\caption{Parameters used in the simulation
\label{tab:rateequations}}
\begin{tabular}{l l l l}
\hline
\textbf{Parameter} & \textbf{Value} & \textbf{Units} & \textbf{Description} \\
\hline
$K_{\mathrm{NV}^-}^{(e)}$ & 27 & $(\mu \mathrm{s})^{-1}\mathrm{/mW}$ & NV$^{-}$ excitation rate \\
$K_{\mathrm{NV}^0}^{(e)}$ & 18 & $(\mu \mathrm{s})^{-1}\mathrm{/mW}$ & NV$^{0}$ excitation rate \\
\hline
$\tau_{\mathrm{NV}^-}$ & $13 \times 10^{-3}$ & $\mu \mathrm{s}$ & Excited-state lifetime (13 ns) \\
$\tau_{\mathrm{NV}^0}$ & $19 \times 10^{-3}$ & $\mu \mathrm{s}$ & Excited-state lifetime (19 ns) \\
\hline
$K_{\mathrm{NV}^-}^{(f)}$ & $1 / \tau_{\mathrm{NV}^-}$ & $(\mu \mathrm{s})^{-1}$ & NV$^{-}$ fluorescence rate \\
$K_{\mathrm{NV}^0}^{(f)}$ & $1 / \tau_{\mathrm{NV}^0}$ & $(\mu \mathrm{s})^{-1}$ & NV$^{0}$ fluorescence rate \\
\hline
$K_{35}$ & 7.9 & $(\mu \mathrm{s})^{-1}$ & NV$^{-}$ ES ms=0 $\to$ singlet \\
$K_{45}$ & 45 & $(\mu \mathrm{s})^{-1}$ & NV$^{-}$ ES ms=$\pm1$ $\to$ singlet \\
$K_{56}$ & 1000 & $(\mu \mathrm{s})^{-1}$ & Singlet ES $\to$ GS \\
$K_{61}$ & 6.5 & $(\mu \mathrm{s})^{-1}$ & Singlet GS $\to$ NV$^{-}$ GS ms=0 \\
$K_{62}$ & 0.1 & $(\mu \mathrm{s})^{-1}$ & Singlet GS $\to$ NV$^{-}$ GS ms=$\pm1$ \\
\hline
$K_{iG}$  & 852  & $(\mu \mathrm{s})^{-1}\mathrm{/mW}$ & Ionization rate (green laser) \\
$K_{rG}$  & 134  & $(\mu \mathrm{s})^{-1}\mathrm{/mW}$ & Recombination rate (green laser) \\
$K_{iIR}$ & 1.2  & $(\mu \mathrm{s})^{-1}\mathrm{/mW}$ & Ionization rate (IR laser) \\
$K_{rIR}$ & 3.17 & $(\mu \mathrm{s})^{-1}\mathrm{/mW}$ & Recombination rate (IR laser) \\
\hline
$\text{$I_G$}$ & 0.7 & mW & Green laser power \\
\hline
\end{tabular}
\end{table*}


\end{document}